\documentstyle[12pt,epsfig]{article}  
\begin{document} 
\baselineskip=18pt

\def\la{\mathrel{\mathpalette\fun <}}
\def\ga{\mathrel{\mathpalette\fun >}}
\def\fun#1#2{\lower3.6pt\vbox{\baselineskip0pt\lineskip.9pt
\ialign{$\mathsurround=0pt#1\hfil##\hfil$\crcr#2\crcr\sim\crcr}}} 

\begin{titlepage} 
\begin{flushright}{\bf Preprint SINP MSU 2004-14/753}\end{flushright}
\begin{center}
$~$ \vskip 5 cm 
{\Large \bf Fast simulation of flow effects in central and semi-central heavy
ion collisions at LHC} \\

\vspace{10mm}

I.P.~Lokhtin and A.M.~Snigirev  \\ ~ \\ 
M.V.Lomonosov Moscow State University, D.V.Skobeltsyn Institute of Nuclear Physics \\
119992, Vorobievy Gory, Moscow, Russia \\ E-mail:~~Igor.Lokhtin@cern.ch   
\end{center}  

\vspace{5mm} 

\begin{abstract} 
The simple method for simulation of ``thermal'' hadron spectra in ultrarelativistic 
heavy ion collisions including longitudinal, transverse and elliptic flow is 
developed. The model is realized as fast Monte-Carlo event generator. 
\end{abstract}

\end{titlepage}   

\section{Introduction} 

The experimental investigation of ultra-relativistic nuclear collisions offers 
a unique possibility of studying the properties of strongly interacting matter 
at high energy density. In that regime, hadronic matter is expected to become 
deconfined, and a gas of asymptotically free quarks and gluons is formed, the 
so-called quark-gluon plasma (QGP), in which the colour interactions between 
partons are screened owing to collective effects~\cite{qm}. One of the 
important tools to study QGP properties is transverse and elliptic flow 
observables. 

In particular, the experimentally observed growth of the mean transverse momentum
with increasing mass in most central nuclear collisions at
SPS~\cite{na44,na49-tr} and RHIC~\cite{star-tr,phenix-tr} energies is
naturally and simply explained with the hydrodynamical 
model~\cite{Rozental:1993}, where the change of momentum $\Delta p_T(\bf r)$ of 
a hadron of mass $m$ due to transverse motion of a fluid element at the point 
$\bf r$ can be written as $\Delta p_T({\bf r})=m \sinh{Y_T({\bf r})}$, where 
$Y_T$ is the collective transverse rapidity. 

Moreover, a strong interest in azimuthal correlation measurements in 
ultrarelativistic heavy ion collisions has recently gained 
impetus. Recent anisotropic flow data from SPS~\cite{na49-el} and 
RHIC~\cite{star-el,phenix-el,phobos} can be described well by hydrodynamical 
models for semi-central collisions and transverse momentum, $p_T$, up to 
$\sim 2$ GeV/c (the coefficient of elliptic flow $v_2$, which is defined as the
coefficient of the second harmonic of the particle azimuthal distribution with 
respect to the reaction plane, appears to be monotonously growing with 
increasing $p_T$~\cite{Kolb:2000} in this case). On the other hand, the 
majority of microscopical Monte-Carlo models underestimate the flow 
effects (see however \cite{Zabrodin:2001}). Measurements of $v_2$ present one 
of the effective tools to test various models. In particular, the saturation 
and gradual decrease of $v_2$ for $p_T > 2$ GeV/c, predicted as a signature of 
strong partonic energy loss in a QGP, is supported by the recent RHIC 
data~\cite{star-el,phenix-el} extending up to $p_T \simeq 10$ GeV/c. The 
interpolation between the low-$p_T$ relativistic hydrodynamics region and 
the high-$p_T$ pQCD-computable region was evaluated in~\cite{Gyulassy:2001}. 

The initial gluon densities in Pb$-$Pb reactions at $\sqrt{s_{\rm NN}}=5.5$~TeV 
at the Large Hadron Collider (LHC) are expected to be significantly higher than 
at RHIC, implying much stronger QGP effects. 
The probing experimental capabilities of LHC detectors together with physics 
and software validation of various Monte-Carlo tools and cross-comparisons 
among different codes are important tasks in the light of coming LHC Heavy Ion 
Program~\cite{mcwshop}. A number of Monte-Carlo generators is available at the 
moment to generate heavy ion events at LHC energies: HIJING~\cite{hijing}, 
FRITIOF~\cite{fritiof}, LUCIAE~\cite{luciae}, DPMJET-III~\cite{dpmjet}, 
PSM~\cite{psm}, NEXUS~\cite{nexus}, etc. However, flow effects in almost of 
such models are lacking or implemented insufficiently. Besides, running these 
codes at LHC energies consumes much computing efforts. On the other hand, 
macroscopic hydrodynamical models basically reproduce the bulk of hadron 
spectra observed at SPS and RHIC, and can be in principle used to estimation of 
particle flow effects at LHC, may be extending this approach to even some 
higher p$_T$ values. Of course, for more detailed simulation, one has to take 
into account the interplay between hydro flow and semi-hard particle flow due 
to parton energy loss, secondary scatterings, etc. 

\section{``Thermal'' model and fast Monte-Carlo generation} 

We suggest simple hydrodynamical Monte-Carlo 
code~\cite{Kruglov:1997,Lokhtin:2002} giving final hadron 
spectrum as a superposition of a thermal distribution and a collective flow  
~\cite{Heinz:1993,Muroya:1995,Lokhtin:1996,Kolb:2000},
\begin{eqnarray}
\label{spectr} 
E\frac{d^3N}{d^3p} = \int_{\sigma}f(x,p)~p^{\mu} d\sigma_{\mu}~~,   
\end{eqnarray} 
where the invariant distribution function $f(x, p)$ is taken in the 
Bose-Einstein form for particles of integer spin and in the Fermi-Dirac form
for particles of half-integer spin ($p_{\mu}$ is the $4$-momentum of hadron, 
and $E=p_0$ is its energy). Integration is performed over the hypersurface 
$\sigma$ at the ``freeze-out'' temperature $T=T_f$. 
The formation of the cylindrically symmetric hot matter expanding preferably 
along the cylinder axis is expected in the case of relativistic heavy ion 
collisions; as to the transverse motion, it can be taken into account as a
correction~\cite{Bjorken:1983}. In this case, the variables 
$\tau$, $r$, $\eta$ and $\Phi$ ($r = \sqrt{x^2 + y^2}, ~~\tau = 
\sqrt{t^2 - z^2}, ~~\eta = \frac{1}{2}\ln{\frac{t+z}{t-z}}, ~~\tan{\Phi}=y/x$) 
are commonly used instead of the Cartesian coordinates $t$, $x$, $y$, $z$. 
We consider charged and neutral pions, kaons and 
nucleons only, and kaons and nucleons are supposed to be thermally suppressed 
by their heavier mass. In addition, the linear transverse velocity profile 
specification 
$$u^r = \sinh{Y_T} = \frac{dR}{d\tau}\frac{r}{R}$$
(which follows from a solution of the nonrelativistic continuity equation 
with uniform density) and longitudinal velocity specification in accordance 
with one-dimensional scaling solution $Y_L = \eta$ are assumed 
($Y_T$ and $Y_L$ are, respectively, the transverse and the longitudinal 
rapidity of collective motion, while $R$ is the effective transverse radius of 
the system).  

The following procedure were applied to simulate ``thermal'' hadron spectra in 
heavy ion AA collisions at given impact parameter $b$. \\
1. The 4-momentum $p^*_{\mu}$ of a hadron of mass $m$ was generated at random 
in the rest frame of a liquid element in accordance with the isotropic 
Boltzmann distribution 
\begin{eqnarray}
f(E^*) \propto E^*\sqrt{E^{*2}-m^2}\exp{(-E^*/T_f)}, ~~~~-1 < \cos{\theta^*} < 
1, ~~~~ 0 < \phi^* < 2\pi ~~, 
\end{eqnarray}  
where $E^*=\sqrt{p^{*2}+m^2}$ is the energy of the hadron, and the polar angle 
$\theta^*$ and the azimuthal angle $\phi^*$ specify the direction of its motion
in the rest frame of the liquid element. \\  
2. The spatial position of a liquid element and its local 4-velocity 
$u_{\mu}$ were generated at random in
accordance with phase space and the character of motion of the fluid:  
\begin{eqnarray}
\label{space}
&     & f(r) = 2r/R_f^2 ~(0 < r < R_f), ~~~~-\eta_{\rm max} < 
\eta <  \eta_{\rm max}, ~~~~
0 < \Phi < 2\pi, \nonumber \\ 
&     & u_r = \frac{r}{R_f}\sinh{Y_T^{\rm max}}, ~~~~u_t=\sqrt{1+u_r^2}\cosh{\eta}, ~~~~ 
u_z=\sqrt{1+u_r^2}\sinh{\eta} ~~,    
\end{eqnarray}  
where $R_f$ is the effective final transverse radius of the system, which is
fixed here by specifying the mean charged multiplicity per unit rapidity 
interval, $<dN/dy^h>$, in the final state; 
$\eta_{\rm max}=Y_L^{\rm max}$ and $Y_T^{\rm max}$ are maximum longitudinal 
and transverse collective rapidities. \\   
3. Further, boost of the hadron 4-momentum in the 
c.m. frame of the event was performed: 
\begin{eqnarray}    
p_x & = & p^*\sin{\theta^*}\cos{\phi^*} + u_r\cos{\Phi}\left[ E^* + \frac{(u^i
p^{*i})}{u_t + 1}\right] \nonumber \\ 
p_y & = & p^*\sin{\theta^*}\sin{\phi^*} + u_r\sin{\Phi}\left[ E^* + \frac{(u^i
p^{*i})}{u_t + 1}\right] \nonumber \\ 
p_z & = & p^*\cos{\theta^*} + u_z\left[ E^* + \frac{(u^i 
p^{*i})}{u_t + 1}\right] \nonumber \\ 
E & = & E^*u_t + (u^i p^{*i}),    
\end{eqnarray} 
where 
\begin{eqnarray}
(u^i p^{*i}) & = & u_rp^*\sin{\theta^*}\cos{(\Phi-\phi^*)} + u_z p^*
\cos{\theta^*} ~.     
\end{eqnarray} 
Anisotropic flow is introduced here under simple assumption that the spatial 
ellipticity of ``freeze-out'' region, 
$\epsilon =\left< y^2 - x^2 \right> / \left< y^2 + x^2 \right> $,
is directly related to the ellipticity of the system formed in the region of the
initial overlap of nuclei, 
$\epsilon _0 = b/2R_A$ ($R_A$ is nucleus radius).  
This ``scaling'' enables one to avoid introducing additional parameters
and, at the same time, leads to an azimuthal anisotropy of generated particles 
due to dependence of effective final radius $R_f (b)$ on the angle  
$\Phi$~\cite{Lokhtin:2000}: 
\begin{equation}
\label{R_f} 
R_f(b) = R_{f}(b=0)~{\rm min}\{ \sqrt{1 - \epsilon^2_0~ \sin^2 \Phi} + \epsilon_0~ 
\cos \Phi , ~~ \sqrt{1 - \epsilon^2_0~ \sin^2 \Phi} - \epsilon_0~ 
\cos \Phi \} . 
\end{equation} 
Obtained in such a way azimuthal distribution of particles is described well by 
the elliptic form for the domain of reasonable impact parameter values.  

We also set the Poisson multiplicity distribution and assume that the mean 
multiplicity of particles is proportional to the 
nuclear overlap function~\cite{Lokhtin:2000}. For estimated ``freeze-out'' 
parameters -- temperature $T_f = 140$ MeV, collective  longitudinal rapidity 
$Y_L^{max}=5$ and collective transverse rapidity $Y_T^{max}=1$ -- we get 
average hadron transverse momentum $<p_T^h> = 0.55$ GeV/c and following
particle ratios: 
$$\pi^{\pm}:K^{\pm}:p^{\pm} = 24:6:1,~~~~~\pi^{\pm}:\pi^0=2:1,
~~~~~K^{\pm}:K^0=1:1,~~~~~p:n=1:1~~.$$   

The model has been realized as fast Monte-Carlo event generator, and
corresponding Fortran routine is available by the web~\cite{hydro}. The following 
input parameters should be specified by user to set hadron event configuration: 
beam and target nucleus atomic number; type of event centrality generation 
(options ``fixed impact parameter'' or ``impact parameter is generated with 
standard Glauber geometry between minimum and maximum values'' are foreseen);  
baseline mean charged particle multiplicity per unit rapidity at mid-rapidity, 
$<dN^{\pm}/dy^h> (y^h=0)$, in central Pb-Pb collisions (total multiplicity for 
other centralities and atomic numbers will be calculated automatically).  
Since the output particle information is stored in common block LUJETS of 
JETSET routine~\cite{pythia}, main users program should be compiled with JETSET 
Fortran routine with extended size (up to 150000) of LUJETS arrays.  

\section{Conclusions} 

The simple model to simulate flow effects in heavy ion collisions at LHC
energies has been developed. This model is realized as fast Monte-Carlo event 
generator, and corresponding Fortran routine is available by the web. \\ 
     
To conclude, let us to discuss the physics validity of the model application. 
\begin{itemize} 
\item Internal parameters of the routine for flow were selected as an estimation 
for LHC heavy ion beam energies. The result for other beam energy ranges, 
obtained without additional internal parameters adjusting, is not expected to 
be reasonable. 
\item Hydro-type description of heavy ion collisions is expected to be 
applicable for central and semi-central collisions. The result obtained for 
very peripheral collisions ($b \sim 2 R_A$) can be not adequate. 
\item Hydro flow mechanism in heavy ion collisions is valid for restricted 
kinematic range: mid-rapidity, low and intermediate $p_T$. The model is not 
applicable for very forward rapidity ($|y| \ga 3$) and very high $p_T$  
($\gg 2-5$ GeV/c).  
\end{itemize} 

{\it Acknowledgments.} \\
We would like to thank S.V.~Petrushanko, C.~Roland and L.I.~Sarycheva for 
useful discussions.

\end{document}